\documentclass[nofootinbib,showpacs,amsmath,amssymb,prd]{revtex4}
\usepackage{graphicx}

\begin{document}

\title{Scale dependence of the quark masses and mixings: leading order}

\author{S.R.~Ju\'{a}rez~W.}
\email{rebeca@esfm.ipn.mx}
\altaffiliation{On sabbatical leave at Departamento de F\'{\i}sica,
CINVESTAV.}

\author{ S.F.~Herrera H.}
\email{simon@esfm.ipn.mx}

\affiliation{Departamento de F\'{\i}sica, Escuela Superior de
F\'{\i}sica y Matem\'{a}ticas, IPN, Mexico}

\author{P.~Kielanowski}
\email{kiel@physics.utexas.edu}

\author{G.~Mora}
\email{Gerardo.Mora@fis.cinvestav.mx}
\altaffiliation{Permanent address:
DACB, Universidad ``Ju\'arez'' Aut\'onoma de Tabasco, Mexico.}

\affiliation{Departamento de F\'{\i}sica, Centro de Investigaci\'{o}n
y Estudios Avanzados, Mexico}

\begin{abstract}
We consider the Renormalization Group Equations (RGE) for the
couplings of the Standard Model and its extensions. Using the
hierarchy of the quark masses and of the Cabibbo-Kobayashi-Maskawa
(CKM) matrix our argument is that a consistent approximation for the
RGE should be based on the parameter $\lambda= |\hat{V}_{us}|
\approx0.22$. We consider the RGE in the approximation where we
neglect all the relative terms of the order $\sim\lambda^{4}$ and
higher. Within this approximation we find the exact solution of the
evolution equations of the quark Yukawa couplings and of the vacuum
expectation value of the Higgs field. Then we derive the evolution of
the observables: quark masses, CKM matrix, Jarlskog invariant,
Wolfenstein parameters of the CKM matrix and the unitarity
triangle. We show that the angles of the unitarity triangle remain
constant. This property may restrict the possibility of new symmetries
or textures at the grand unification scale.
\end{abstract}


\maketitle

\section{Introduction}
\label{section_I}
The solutions of the Renormalization Group Equations (RGE) determine
the dependence of the physical parameters of the theory on the
renormalization
point~\cite{ref1,ref2,ref3,ref4,ref7,ref5,n2,n3,n4,n6,n7}. The set of
these parameters for a given energy defines the theory that is
equivalent to the one with the initial parameters. This property was
used in the Standard Model in the search of the theory that is
equivalent to the low energy Standard Model. This led to the
hypothesis of grand unification when it was observed that 3~gauge
coupling constants of the Standard Model converge to one value at the
energy $\sim10^{14}\,$GeV, and to the reduction of the number
of the parameters of the Standard Model. Further reduction of the
number of the parameters is possible in the sector of the
Cabibbo-Kobayashi-Maskawa (CKM) matrix~\cite{n9,n10} by looking for
the textures or new symmetries. The CKM matrix appears in the Standard
Model as a result of the transition from the quark gauge eigenstates
to the quark mass eigenstates upon the diagonalization of the quark
mass matrices. The quark mass matrices appear after the spontaneous
symmetry breaking from the quark-Higgs Yukawa couplings. For this
reason we consider the RGE for the quark-Higgs Yukawa couplings from
which we obtain the evolution of the quark masses and the CKM~matrix.

In a recent paper~\cite{ref0} we began the analysis of the solutions
of the one and two-loop evolution equations (obtained in a general
quantum field theory) for the coupling constants, the quark Yukawa
couplings (QYC) and the CKM~matrix.  In~Ref.~\cite{ref0} we
systematically investigated the influence of the hierarchical
structure of the QYC on the evolution of the CKM~matrix by
constructing the exact solution of the one loop RGE compatible with
the observed hierarchy The most important result that we derived is
that the CKM~matrix evolution depends only on one universal function
of energy which is a suitable integral that depends on the model
(Standard Model~(SM), Minimal Supersymmetric Standard Model~(MSSM) and
Double Higgs Model~(DHM)). We showed that the evolution of the ratios
of the eigenvalues (masses) of the \textit{up} and \textit{down} QYC,
$Y_{u,c}/Y_{t}$ and $Y_{d,s}/Y_{b}$ depend on the same universal
function as the CKM~matrix. The eigenvalues of the QYC, $Y_{u}$ and
$Y_{c}$ depend linearly on the corresponding initial values and their
ratio $Y_{u}/Y_{c}$ is constant while the functional dependence for
$Y_{t}$ is non linear. The other remarkable result is that the
diagonalizing matrices of the \textit{up} quark Yukawa couplings are
energy independent in the leading order. This means that the
transformation $(\psi _{u})_{L,R}\rightarrow (U_{u})_{L,R}(\psi
_{u})_{L,R}$, will diagonalize the matrix of the \textit{up} quark
Yukawa couplings and it will stay diagonal upon the renormalization
group evolution, and the evolution of the CKM~matrix will be
determined only from the evolution of the \textit{down} quarks Yukawa
couplings. This fact may simplify the model building based on the
symmetries of the quark Yukawa couplings.

To achieve our present goal to obtain the precise results for the
running of the parameters of the standard model and its extensions we
will use the hierarchy and approximation scheme based on the parameter
$\lambda=|\hat{V}_{us}|\approx0.22$ and establish accordingly the
consistent procedure for the solution of the RGEs. We show that a one
loop approximation is equivalent to neglecting in RGEs of the terms of
the \emph{relative} order $\lambda^{4}$ and higher. This means that
the terms of this type should also be neglected, if present, in the
one loop RGEs. The inclusion of the terms of the relative order
$\lambda^{4}$ and $\lambda^{5}$ requires the two loop part of the RGEs
and for the terms of the order $\lambda^{8}$ one has to include the
three loop terms.

Our aim is to systematically investigate the running of the
observables of the currently available models up to the order
$\lambda^{5}$ which we plan to do in two steps. In the first step we
consider the RGEs up to the order $\lambda^{3}$ and use the fact that
these equations can be analytically solved. In the second step we
apply the modified perturbation calculus to obtain the evolution of
the parameters with the precision $\lambda^{5}$. The methods and scope
of each step are very wide apart so we will divide our analysis into
two papers. In the present paper we establish the framework of our
analysis, present exact solutions of the RGEs for the Yukawa
couplings, vacuum expectation values and other observables and also
show a graphical representation of the analytical results. In the
forthcoming paper, using these results, we will develop the modified
perturbation calculus which will be the basis of the derivation of the
corrections up to the order $\lambda^{5}$ to the analytical results
from the first paper.

The organization of the paper is the following. The
Section~\ref{section_IIa} is devoted to the discussion of the
hierarchy of the couplings and the observables in the RGEs and the
introduction of the approximation scheme that we apply to the
solution.  In Section~\ref{section_II} we discuss RGEs for the Yukawa
couplings and the vacuum expectation values in the $\lambda^{3}$ order
and present the analytical solutions. The behavior of the universal
energy function $h(t)$ and of all the other functions that participate
in the evolution of the physical parameters of the SM and its
extensions is presented graphically from the mass scale $m_{t}$ to the
Planck mass.  In Section~\ref{section_III}, with the exact solutions
of the one loop RGE equations previously obtained, we solve,
considering a new approach, the equations for the other observables
(all the CKM~matrix elements, the masses, unitarity triangle, the
Jarlskog invariant $J$ and the Wolfentstein parameters $\lambda$, $A$,
$\rho$, $\eta$), we present their explicit analytical evolution which
were not published before. Subsequently we represent graphically the
RGE flow of these observables.  In Section~\ref{section_IV} we draw
the conclusions. Our new approach confirms the earlier results and
also enables us to obtain the new ones, such as the differential
equations for the diagonalizing matrices of the down sector in terms
of which the CKM matrix elements are obtained.

\section{Approximation scheme}
\label{section_IIa}

In the approximate calculations the overall consistency is very
important. This means that the terms that are smaller than those
neglected should not be kept. The order of magnitude (expressed in the
powers of $\lambda$) of the components of the RGEs is
\begin{equation}
\label{H1}
\begin{split}
&g_{1}^2\approx 0.2 \sim\lambda,\;\;\;
g_{2}^2\approx 0.42 \sim2\lambda,\;\;\;
g_{3}^2\approx 1.5 \sim 1,\;\;\;
y_{u}y_{u}^{\dagger} \sim 1,\\
&y_{d}y_{d}^{\dagger} \sim \left(\frac{m_{b}}{m_{t}}\right)^2\approx
1.13\lambda^{5},\;\;\;
\frac{1}{(4\pi)^{2}}=6.3\cdot 10^{-3}\approx
2.7\lambda^4,\;\;\;
\lambda_{H}\approx 0.44\sim2\lambda.\\
\end{split}
\end{equation}
Here $g_{i}$ are the gauge coupling constants, $y_{u}$ and $y_{d}$
are the matrices of the Yukawa couplings and $\lambda_{H}$ is the
Higgs quartic coupling. 

The mathematical structure of the RGEs is the following
\begin{subequations}
\label{RGEE}
\begin{eqnarray}
&&
\label{RGEEa}
\frac{dg_{i}}{dt}=
\frac{1}{(4\pi)^{2}}
\left\{
{\beta}^{(1)}_{g_{i}}(g_{i}^{2})+
\frac{1}{(4\pi)^{2}}
{\beta}^{(2)}_{g_{i}}(g_{k}^{2},y_{u}y_{u}^{\dagger},
y_{d}y_{d}^{\dagger})+\cdots
\right\}g_{i}\\
&&
\label{RGEEb}
\frac{dy_{u,d}}{dt}=
\frac{1}{(4\pi)^{2}}
\left\{
{\beta}^{(1)}_{u,d}(g_{k}^{2},y_{u}y_{u}^{\dagger},
y_{d}y_{d}^{\dagger})
+\frac{1}{(4\pi)^{2}}
{\beta}^{(2)}_{u,d}(g_{k}^{2},y_{u}y_{u}^{\dagger},
y_{d}y_{d}^{\dagger},\lambda_{H})+\cdots
\right\}y_{u,d}\\
&&
\label{RGEEc}
\frac{dv}{dt}=
\frac{1}{(4\pi)^{2}}
\left\{
{\beta}^{(1)}_{v}(g_{k}^{2},
y_{u}y_{u}^{\dagger},
y_{d}y_{d}^{\dagger})
+\frac{1}{(4\pi)^{2}}
{\beta}^{(2)}_{v}(g_{k}^{2},
y_{u}y_{u}^{\dagger},
y_{d}y_{d}^{\dagger},\lambda_{H})+\cdots
\right\}v\\
&&
\label{RGEEd}
\frac{d\lambda_{H}}{dt}=
\frac{1}{(4\pi)^{2}}
\left\{
{\beta}^{(1)}_{\lambda_{H}}(g_{k}^{2},
y_{u}y_{u}^{\dagger},
y_{d}y_{d}^{\dagger})
+\frac{1}{(4\pi)^{2}}
{\beta}^{(2)}_{\lambda_{H}}(g_{k}^{2},
y_{u}y_{u}^{\dagger},
y_{d}y_{d}^{\dagger},\lambda_{H})+\cdots
\right\}
\end{eqnarray}
\end{subequations}
In Eqs.~(\ref{RGEE}) 
$t\equiv \ln (E/\mu )$ is the energy scale parameter, $\mu$ is the
renormalization point, $v$ is the the
vacuum expectation value of the Higgs field,
$\beta^{(n)}$ denote the $n$-loop $\beta$ functions for the RGEs which are
homogenous polynomials of the indicated variables and the dots indicate the
omitted three and more loops contribution. The approximation scheme
for the RGEs consists in the approximation of the terms inside the braces
on the right hand side of the equations keeping only the terms of
a given relative order.

From Eqs.~(\ref{RGEE}) and~(\ref{H1}) one can see  that the leading
terms of the one loop contribution are of the order~1, because
\begin{equation}
\label{H2}
{\beta}^{(1)}(g_{k}^{2},\ldots)\sim 1
\end{equation}
and for the two loop contribution
\begin{equation}
\label{H3}
\frac{1}{(4\pi)^2}{\beta}^{(2)}(g_{k}^{2},\ldots)\sim
\frac{1}{(4\pi)^2}\sim\lambda^{4}. 
\end{equation}
From Eqs.~(\ref{H2}) and~(\ref{H3}) we thus see that in the one loop
approximation we should neglect all the terms of the order
$\lambda^{4}$ and higher. From the hierarchy given in Eq.~(\ref{H1})
it follows that the one loop approximation is equivalent to the
neglecting the two loop term and also putting $y_{d}y_{d}^{\dagger}=0$
on the right hand side of~(\ref{RGEE}) \emph{relatively} to the terms
of the lower order. Such equations, as it turns out, can be
analytically solved.

The next order term is of the order $\lambda^{4}$. Such a term is
present only in the two loop contribution and it does not introduce
important qualitative modification to the running of the observables
and it will be discussed together with Eq.~(\ref{RGEEd}) and the
$\lambda^{5}$ contribution in the forthcoming paper. RGEs for the
order $\lambda^{4}$ and higher form a system of coupled nonlinear
equations which are difficult to solve analytically, so for their
analysis one has to use other methods.

\section{One loop equations and solutions for the gauge coupling parameters,
the Yukawa couplings and the Higgs vacuum expectation value}
\label{section_II}

In this section we discuss the {\em one loop} RGEs for the gauge,
Yukawa couplings and the vacuum expectation values. This means that
the precision of all the expressions is up to the order
$\lambda^{3}$. We indicate this fact in the final form of the equation
for each observable. The structure of the one-loop RGE for the gauge
coupling parameters
$g_{k}$ is
\begin{equation}
\frac{dg_{i}}{dt}=\frac{1}{(4\pi )^{2}}b_{i}g_{i}^{3}+{\cal O}(\lambda^{4}).
\label{P1}
\end{equation}
Here the coefficients $(b_1,b_2,b_3)$ are equal $(41/10,-19/6,-7)$,
$(21/5,-3,-7)$ and $(33/5,1,-3)$ for the SM, DHM and MSSM,
respectively. The solution to this equation is derived directly
\begin{equation}
g_{i}(t)=g_{i}^{0}\left( 1-
\frac{2b_{i}(g_{i}^{0})^{2}(t-t_{0})}{(4\pi )^{2}}\right)^{-1/2},
\;\;\;\;
g_{i}^{0}\equiv g_{i}(t_{0}).  \label{P2}
\end{equation}

The one-loop RGE for the Yukawa couplings $y_{u,d }$ are 
\begin{equation}
\frac{dy_{u,d}}{dt}=\left[ \frac{1}{(4\pi )^{2}}
\beta_{u,d}^{(1)}\right] y_{u,d}+{\cal O}(\lambda^{4}).  \label{P3}
\end{equation}
The $\beta _{u,d}^{(1)}$ has a hierarchical structure based on the
parameter $\lambda$ and has the following form\footnote{
Notice that the form of the function $\beta^{(1)}_{u,d}$ is
compatible with Eq.~(\ref{RGEEb}), i.e.\ it is the homogenous
polynomial of order~1 of the indicated variables.}
(with lepton dependence suppressed)
\begin{equation}
\beta^{(1)}_i=
\alpha^{i}_1(t)+
\alpha^{i}_2(y_{u}^{\vphantom{\dagger}}y_{u}^{\dagger})+
\alpha^{i}_3\text{Tr}(y_{u}^{\vphantom{\dagger}}y_{u}^{\dagger})+
\alpha^{i}_4(y_{d}^{\vphantom{\dagger}}y_{d}^{\dagger})+
\alpha^{i}_5\text{Tr}(y_{d}^{\vphantom{\dagger}}y_{d}^{\dagger}),
\;\;i=u,d.
\label{beta1}
\end{equation}
Its structure allows the systematical solution of the RGE with
$\lambda $ as the expansion coefficient. The approximate form of the
equations for the quark Yukawa couplings, neglecting all the terms of
$\lambda ^{4}$ and higher is the following
\begin{eqnarray}
\frac{dy_{u}}{dt} &=&\frac{1}{(4\pi )^{2}}[\alpha _{1}^{u}(t)+\alpha
_{2}^{u}y_{u}^{\phantom {\dagger }}y_{u}^{\dagger }+\alpha _{3}^{u}
\text{Tr}
(y_{u}^{\phantom {\dagger }}y_{u}^{\dagger })]y_{u}
+{\cal O}(\lambda^{4}),  \label{1} \\
\frac{dy_{d}}{dt} &=&\frac{1}{(4\pi )^{2}}[\alpha _{1}^{d}(t)+\alpha
_{2}^{d}y_{u}^{\phantom {\dagger }}y_{u}^{\dagger }+\alpha _{3}^{d}\text {Tr}
(y_{u}^{\phantom {\dagger }}y_{u}^{\dagger })]y_{d}
+{\cal O}(\lambda^{4}),  \label{2}
\end{eqnarray}
where 
\begin{eqnarray*}
\alpha _{1}^{u}(t) &=&-(c_1g_1^2+c_2g_2^2+c_3g_3^2),
\,\,\alpha _{2}^{u}=\frac{3b}{2}
,\,\,\,\,\alpha _{3}^{u}=3, \\
\alpha _{1}^{d}(t)
&=&-(c_1^{\prime}g_1^2+c_2^{\prime}g_2^2+c_3^{\prime}g_3^2)
,\,\,\,\alpha_{2}^{d}=\frac{3c}{2},\,\,
\alpha _{3}^{d}=3a,
\end{eqnarray*}
$(a,b,c)$ are equal to $(1,1,-1)$,$(0,1,1/3)$, $(0,2,2/3)$;
$(c_1,c_2,c_3)$ are equal to $(17/20,9/4,8)$, $(17/20,9/4,8)$,
$(13/15,3,16/3)$ and $(c_1^{\prime},c_2^{\prime},c_3^{\prime})$ are
equal to $(1/4,9/4,8)$, $(1/4,9/4,8)$, $(7/15,3,16/3)$ in the SM, DHM
and MSSM, respectively.

The transformation from the quark gauge states to the physical states
requires the diagonalization of the Yukawa coupling matrices $y_{u}$
and $y_{d}$ with the biunitary transformations
\begin{equation}
\text{Diag}
(Y_{u(d)},Y_{c(s)},Y_{t(b)})=(U_{u(d)})^{\vphantom{\dagger}}_{L}
y_{u(d)}(U_{u(d)})_{R}^{\dagger }
\label{3}
\end{equation}
where $(U_{u,d})_{L,R}$ are the corresponding unitary diagonalizing
matrices and $Y_{u(d)},Y_{c(s)}$ and $Y_{t(b)}$ are the eigenvalues of the
Yukawa coupling matrices $y_{u(d)}$~\cite{ref0A}.  
As it is well known, the diagonalizing matrices generate the
flavor mixing in the charged current described by the CKM~matrix
\[
\hat{V}=(U_{u})^{\vphantom{\dagger}}_{L}(U_{d})_{L}^{\dagger }. 
\]
The same biunitary diagonalization transformation also permits the
exact solution of Eq.~(\ref{1}) which arises from the fact that the
diagonalizing matrices $(U_{u})_{L,R}$ for the one loop Eq.~(\ref{1})
do not depend on energy~\cite{ref0}. It then follows that
$y_{u}(t)$ from Eq.~(\ref{1}) has the following representation
\begin{equation}
y_{u}(t)=(U_{u})_{L}^{\dagger }Y^{u}(t)(U_{u})_{R}^{\phantom {\dagger }
},\,\,\,Y^{u}(t)=\text{Diag}(Y_{u},Y_{c},Y_{t})  \label{4}
\end{equation}

The whole dependence on $t$ in Eq.~(\ref{4}) is contained only in the
diagonal matrix $Y^{u}(t)$. The diagonal elements of $Y^{u}(t)$
satisfy the system of differential equations that follows from
Eq.~(\ref{1})

\begin{subequations}
\label{5all}
\begin{eqnarray}
\frac{dY_{u,c}}{dt} &=&\frac{1}{(4\pi )^{2}}\{\alpha _{1}^{u}(t)+\alpha
_{3}^{u}Y_{t}^{2}\}Y_{u,c}+{\cal O}(\lambda^{4}),  \label{5a} \\
\frac{dY_{t}}{dt} &=&\frac{1}{(4\pi )^{2}}\{\alpha _{1}^{u}(t)+(\alpha
_{2}^{u}+\alpha _{3}^{u})Y_{t}^{2}\}Y_{t}+{\cal O}(\lambda^{4}).  \label{5c}
\end{eqnarray}
\end{subequations}
Eq.~(\ref{5c}) decouples from Eqs.~(\ref{5a}) and can
be solved independently of the other equations. When the solution for
$Y_{t}(t)$ is known then Eqs.~(\ref{5a}) become
linear and can also be solved and eventually the solution of
Eqs.~(\ref{5all}) is
\begin{subequations}
\label{6all}
\begin{eqnarray}
Y_{u,c}(t) &=&Y_{u,c}(t_{0})\sqrt{r(t)}h_{m}^{\alpha _{3}^{u}}(t)\,,
\label{6a} \\
Y_{t}(t) &=&Y_{t}(t_{0})\sqrt{r(t)}\,h_{m}^{\left( \alpha
_{2}^{u}+\alpha _{3}^{u}\right) }(t).  \label{6c}
\end{eqnarray}
where the functions $r(t)$ and $h_{m}(t)$ are 
\end{subequations}
\begin{eqnarray}
r(t) &=&\exp \left[ \frac{2}{(4\pi )^{2}}\int_{t_{0}}^{t}\alpha
_{1}^{u}(\tau )d\tau \right] =\prod_{k=1}^{k=3}\left[ \frac{g_{k}^{2}\left(
t_{0}\right) }{g_{k}^{2}\left( t\right) }\right] ^{\frac{c_{k}}{b_{k}}},
\label{7a} \\
h_{m}(t) &=&\exp \left(\frac{1}{(4\pi
)^{2}}\int_{t_{0}}^{t}Y_{t}^{2}(\tau )d\tau\right)
=\left( \frac{1}{1-\frac{2\left( \alpha _{2}^{u}+\alpha _{3}^{u}\right) }{
\left( 4\pi \right) ^{2}}(Y_{t}^{0})^{2}\int_{t_{0}}^{t}r(\tau )d\tau }
\right) ^{\frac{1}{2\left( \alpha _{2}^{u}+\alpha _{3}^{u}\right) }}.
\label{7b}
\end{eqnarray}
It is worth mentioning that the function $h(t)\equiv\left[ h_{m}\left(
t\right) \right] ^{\alpha _{2}^{d}}$ appears in
the evolution of the $\hat{V}$ matrix and moreover this is the only
dependence on $t$ of the $\hat{V}$ matrix.

The next step is the determination, from Eq.~(\ref{2}), of the running
of the \textit{down} quark Yukawa couplings. The substitution
\begin{equation}
y_{d}(t)=(U_{u})_{L}^{\dagger }W(t)  \label{8}
\end{equation}
transforms Eq.~(\ref{2}) into the following equation 
\begin{equation}
\frac{dW}{dt}=\frac{1}{(4\pi )^{2}}\left\{ \alpha _{1}^{d}(t)+\alpha
_{2}^{d}\left( Y^{u}(t)\right) ^{2}+\alpha _{3}^{d}\text {Tr}\left(
Y^{u}(t)\right) ^{2}\right\} W,  \label{9}
\end{equation}
with the matrix $\left\{ \alpha _{1}^{d}(t)+\alpha _{2}^{d}\left(
Y^{u}(t)\right) ^{2}+\alpha _{3}^{d}\text
{Tr}\left( Y^{u}(t)\right) ^{2}\right\} $ being diagonal.

This allows to solve Eq.~(\ref{9}) explicitly and the solution for
$W(t)$ reads
\begin{equation}
W(t)=(r^{\prime }(t))^{1/2}h_{m}^{\alpha _{3}^{d}}(t)\,\cdot Z(t)\cdot
W(t_{0})  \label{10}
\end{equation}
where $Z(t)$ is the diagonal matrix 
\begin{equation}
Z(t)=\text{Diag}(1,1,h(t)).  \label{11}
\end{equation}
Now, with the help of Eq.~(\ref{8}) one gets the explicit one loop
running of the Yukawa couplings $y_{d}(t)$
\begin{equation}
y_{d}(t)=\sqrt{r^{\prime }(t)}h_{m}^{\alpha
_{3}^{d}}(t)(U_{u})_{L}^{\dagger }Z(t)(U_{u})_{L}^{\phantom {\dagger }
}y_{d}^{0},\;\;\;\;y_{d}^{0}\equiv y_{d}(t_{0}),
\label{12}
\end{equation}
where 
\begin{equation}
r^{\prime }(t)=\exp \left[ \frac{2}{(4\pi )^{2}}\int_{t_{0}}^{t}\alpha
_{1}^{d}(\tau )d\tau \right] =\prod_{k=1}^{k=3}\left[ \frac{g_{k}^{2}\left(
t_{0}\right) }{g_{k}^{2}\left( t\right) }\right] ^{\frac{c_{k}^{\prime }}{
b_{k}}}.  \label{13}
\end{equation}

The approximated one-loop RGE for the Higgs vacuum expectation value (VEV)
$v_{u,d}\,$ is the following
\begin{equation}
\frac{dv_{u,d}}{dt}=\frac{1}{(4\pi )^{2}}[\alpha
_{1}^{v_{u,d}}(t)+\alpha _{3}^{v_{u,d}}\text {Tr}(y_{u}^{\phantom {
\dagger }}y_{u}^{\dagger })]v_{u,d}+{\cal O}(\lambda^{4}),  \label{15}
\end{equation}
where $v_{u}$ is the VEV for the \textit{up} quarks and $v_{d}$ 
for the \textit{down} quarks\footnote{We consider only the case where
$\tan{\beta}=v_d/v_u\sim 1$. Large values of $\tan{\beta}$ require
other treatment and will be considered elsewhere.}. Note that for the
SM $v_{o}=v_{u}=v_{d}$.
The functions $\alpha_{1}^{v_{u,d}}(t)$ and the coefficients
$\alpha _{3}^{v_{i}}$ are equal
\begin{eqnarray}
&\alpha_1^{v_{u,d}}(t)=c^{\prime\prime}_{1(u,d)}g_1^2+
c^{\prime\prime}_{2(u,d)}g_2^2\\
&\alpha _{3}^{v_{o}}=-3,\;\;\;\;\;
\alpha_{3}^{v_{u}}=-3,\;\;\;\;\;
\alpha _{3}^{v_{d}}=0,
\end{eqnarray}
and the constants
$(c^{\prime\prime}_{1(u,d)},c^{\prime\prime}_{2(u,d)})$ are equal to
$(-9/20,-9/4)$, $(-9/20,-9/4)$ and $(-3/20,-3/4)$ for the SM, DHM and
MSSM, respectively.

\begin{figure}[!tb]
\centering
\includegraphics[angle=90,height=0.9\textheight]{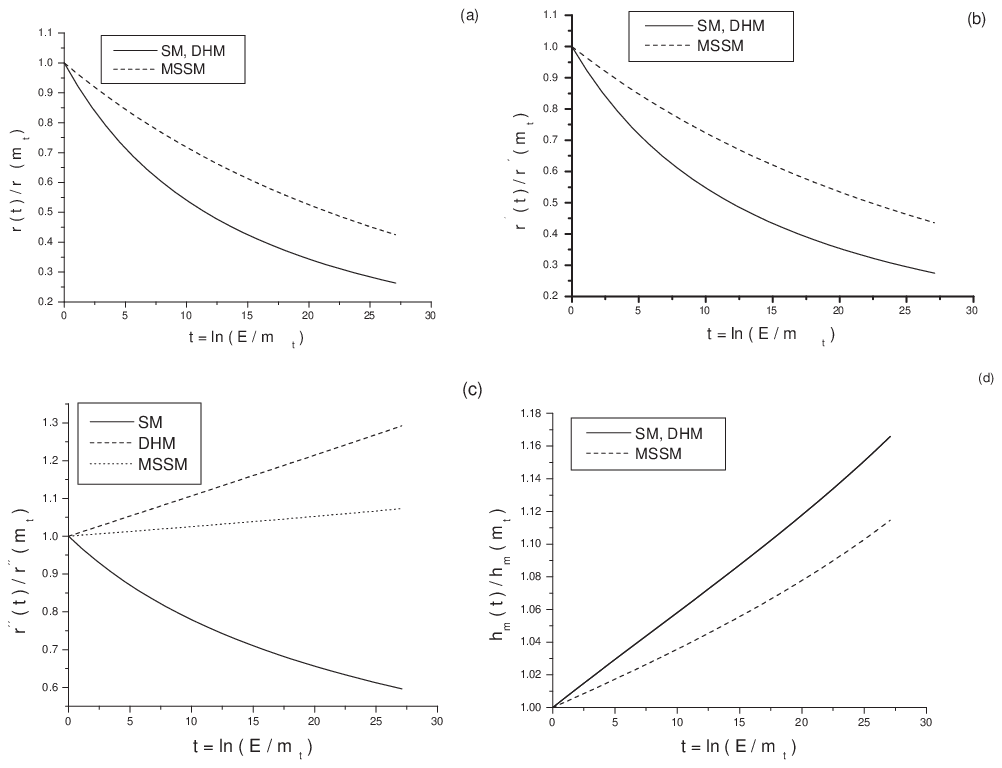}
\caption{\label{fig1}The scale dependence of the ratios:
$r(t)/r(m_{t})$, $r^{\prime}(t)/r^{\prime}(m_{t})$,
$r^{\prime\prime}(t)/r^{\prime\prime}(m_{t})$ and
$h_{m}(t)/h_{m}(m_{t})$.}  
\end{figure}

\begin{figure}[!tb]
\centering
\includegraphics{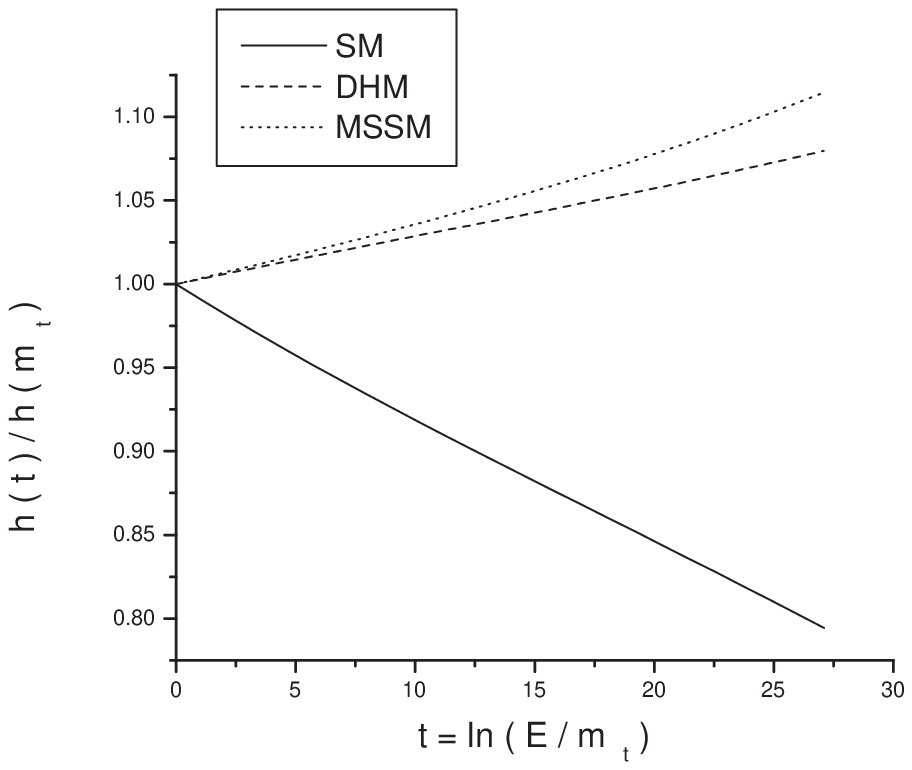}
\caption{\label{fig2}The scale dependence of the ratio $h(t)/h(m_{t})$.}
\end{figure}

To solve Eq.~(\ref{15}) for the VEV  we divide both sides of the equation
by the corresponding VEV and obtain after the integration
\begin{equation}
\int_{t_{0}}^{t}\frac{dv_{u,d}}{v_{u,d}}=\frac{1}{(4\pi )^{2}}
\int_{t_{0}}^{t}[\alpha _{1}^{v_{u,d}}(t)+\alpha _{3}^{v_{u,d}}\text {Tr}
(y_{u}^{\phantom {\dagger }}y_{u}^{\dagger })]dt.  \label{18}
\end{equation}
The left hand side of Eq.~(\ref{18}) is equal to $\ln \left[ v_{u,d}\left(
t\right) /v_{u,d}\left( t_{0}\right) \right] $ and on the right hand side
we have the sum of two integrals. The first integral can be explicitly
integrated using Eq.~(\ref{P2}) and we introduce the function $
r^{\prime\prime}_{{v_{u,d}}}(t)$ 
\begin{equation}
r^{\prime\prime}_{{v_{u,d}}}(t)=\exp \left[ \frac{2}{(4\pi
)^{2}}\int_{t_{0}}^{t}
\alpha _{1}^{v_{u,d}}(\tau )d\tau \right] =\prod_{k=1}^{k=2}\left[ \frac{
g_{k}^{2}(t_{0})}{g_{k}^{2}(t)}\right] ^{\frac{c_{k(u,d)}^{\prime \prime }
}{b_{k}}}  \label{19}
\end{equation}
and the second integral is calculated using Eq.~(\ref{7b}) 
\begin{equation}
\frac{\alpha _{3}^{v_{u,d}}}{(4\pi )^{2}}\int_{t_{0}}^{t}\text {Tr}(y_{u}^{
\phantom {\dagger }}y_{u}^{\dagger })d\tau =\frac{\alpha _{3}^{v_{u,d}}}{
(4\pi )^{2}}\int_{t_{0}}^{t}Y_{t}^{2}(\tau )d\tau =\alpha
_{3}^{v_{u,d}}\ln h_{m}(t).  \label{20}
\end{equation}
The final form of the VEV is thus equal to
\begin{equation}
v_{u,d}\left( t\right) =v_{u,d}\left( t_{0}\right) \sqrt{
r^{\prime\prime}_{v_{u,d}}(t)}h_{m}^{\alpha
_{3}^{v_{u,d}}}(t),\,\,\,\,\,\,\,\,\,\,\,\,\,\,\,\,
\,v_{u,d}^{0}=v_{u,d}\left( t_{0}\right).  \label{21}
\end{equation}
The solutions of the renormalization group equations for the gauge coupling
constants, Eq.~(\ref{P2}), for the quark Yukawa couplings
Eqs.~(\ref{6all}), (\ref{12}) and~(\ref{II.12}) and for the vacuum
expectation values
Eq.~(\ref{21}) 
form the complete
set of the evolution functions from which one can obtain the renormalization
group flow of all observables related to quarks: \textit{up} and
\textit{down} quark
masses and the CKM~matrix. In the next sections we will analyze these
observables but let us notice here that the evolution is described by the
following functions of energy 
\begin{equation}
r(t),r^{\prime }(t),
r^{\prime\prime}_{v_{u,d}}(t),h_{m}(t),h(t).  \label{22}
\end{equation}
The dependence of the observables on these functions will
be discussed in the next sections. Here in Figs.~\ref{fig1}
and~\ref{fig2} we show the functional dependence of
the functions in Eq.~(\ref{22}) for three models (SM, DHM and MSSM) to be
able to see what is their influence on the
observables and how they depend on the model. Notice that in all the
figures we choose
as the renormalization point the mass of the top quark
$m_t=174.3$~GeV~\cite{PDG}. In such 
a way the functions in Eq.~(\ref{22}) and
observables are independent of the quark mass thresholds. The
extensive discussion of the thresholds effects is given in Ref.~\cite{ref4}.

\section{Evolution of the observables}
\label{section_III}

The solution of the renormalization group equations presented in
Section~\ref{section_II} allows the analysis of the evolution of all
the observables related with the Yukawa couplings and the Higgs vacuum
expectation values, i.e.~the quark masses and the
Cabibbo-Kobayashi-Maskawa matrix. The results of this section have been
obtained from the explicit solutions from the previous section. Their
validity and precision are therefore the same, i.e.~the terms of the
order $\lambda^{4}$ and higher have been neglected.  We start with the
analysis of the quark masses presenting first the analytical results
and then showing the corresponding graphs. The same type of the
analysis will be also applied to the CKM~matrix.

\subsection{Quark masses}
\label{section_III_A}

The quark masses after the spontaneous symmetry breaking are equal to
\begin{equation}
m_{i}=\frac{{v}_{i}}{\sqrt{2}}Y_{i} \label{II.1}
\end{equation}
where $Y_{i}$ are the eigenvalues of the corresponding Yukawa
couplings and ${v}_{i}$ is the vacuum expectation value of the Higgs
field. For the theories with one Higgs doublet (SM) there is one Higgs
vacuum expectation value and for two Higgs doublets (DHM, MSSM) there
is one VEV for the \textit{up} quarks and another for the
\textit{down} quarks.

\subsubsection{Up quark masses}
\label{section_III_A_1}

\begin{figure}[!tb]
\centering
\includegraphics[angle=90,height=0.9\textheight]{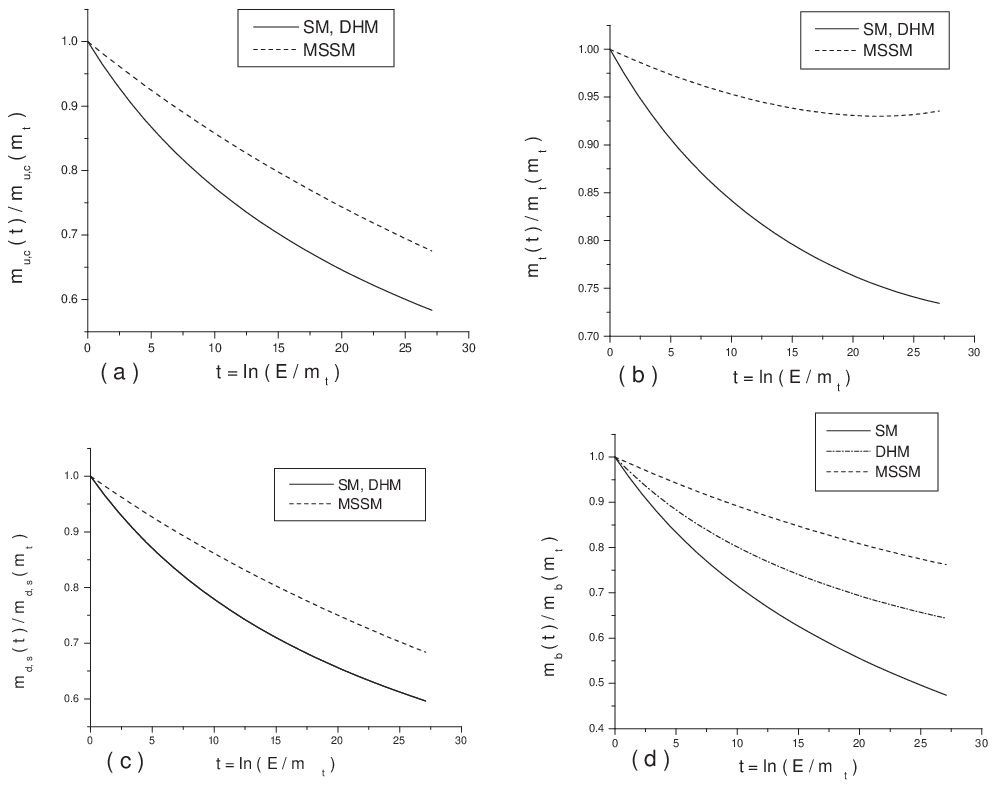}
\caption{\label{fig3}The scale dependence of the ratios
$m_i(t)/m_i(m_{t})$ for the \textit{up} and \textit{down} quarks.}
\end{figure}

The evolution of the eigenvalues $\left( Y_{u},Y_{c},Y_{t}\right) $ for the
\textit{up} quarks is given in Eqs.~(\ref{6all}) and the
evolution of the
VEV's is given in Eq.~(\ref{21}). Using Eq.~(\ref{II.1}) we thus obtain 
\begin{equation}
m_{i}^{u}\left( t\right) =\frac{{v}_{u}(t)}{\sqrt{2}}Y_{i}^{u}(t)=\frac{{v}
_{u}(t_{0})}{\sqrt{2}}Y_{i}^{u}(t_{0})\sqrt{r(t)
r^{\prime\prime}_{{v}_{u}}(t)}
h_{m}^{K_{i}^{u}}\left( t\right)
=m_{i}^{u}(t_{0})
\sqrt{r(t)r^{\prime\prime}_{{v}_{u}}(t)}
h_{m}^{K_{i}^{u}}\left(t\right),  \label{II.2}
\end{equation}
and the power $K_{i}^{u}$ is equal 
\begin{equation}
K_{u,c}^{u}=\alpha _{3}^{u}+\alpha _{3}^{{v}_{u}},\,\,\,\,\,\,\,K_{t}^{u}=
\alpha _{2}^{u}+\alpha _{3}^{u}+\alpha _{3}^{{v}_{u}}.  \label{II.3}
\end{equation}
In Figs.~\ref{fig3}~(a) and~\ref{fig3}~(b) we show the running of the
ratios of
the \textit{up} quark masses  $m_{i}(t) /m_{i}(t_{0})$ $i=u, c, t$ for three
models SM, DHM and MSSM.

\subsubsection{Down quark masses}
\label{section_III_A_2}

The evolution of the eigenvalues of the \textit{down} quark Yukawa couplings is more
complicated because the diagonalizing matrices are also running. If we write
the matrix $y_{d}$ in the form 
\begin{equation}
y_{d}=\left( U_{L}^{d}\right) ^{\dagger }Y^{d}\left( U_{R}^{d}\right),
\label{II.4}
\end{equation}
then Eq.~(\ref{2}) becomes 
\begin{equation}
\frac{d\left( \left( U_{L}^{d}\right) ^{\dagger }Y^{d}\left(
U_{R}^{d}\right) \right) }{dt}=\frac{1}{(4\pi )^{2}}\left\{ \alpha
_{1}^{d}(t)+\alpha _{2}^{d}y_{u}^{\phantom {\dagger }}y_{u}^{\dagger
}+\alpha _{3}^{d}\text {Tr}(y_{u}^{\phantom {\dagger }}y_{u}^{\dagger
})\right\} \left( \left( U_{L}^{d}\right) ^{\dagger }Y^{d}\left(
U_{R}^{d}\right) \right),  \label{II.5}
\end{equation}
which after some simple manipulations becomes 
\begin{equation}
U_{L}^{d}\frac{d\left( U_{L}^{d}\right) ^{\dagger }}{dt}Y^{d}+\frac{dY^{d}
}{dt}+Y^{d}\frac{d\left( U_{R}^{d}\right) }{dt}\left( U_{R}^{d}\right)
^{\dagger }
=\frac{1}{(4\pi )^{2}}\left( U_{L}^{d}\right) \left\{ \alpha
_{1}^{d}(t)+\alpha _{2}^{d}y_{u}^{\phantom {\dagger }}y_{u}^{\dagger
}+\alpha _{3}^{d}\text {Tr}(y_{u}^{\phantom {\dagger }}y_{u}^{\dagger
})\right\} \left( U_{L}^{d}\right) ^{\dagger }Y^{d}
\equiv\frac{1}{(4\pi )^{2}}M^{{vv}}Y^{d},  \label{II.6}
\end{equation}
\begin{equation}
M^{{vv}}\equiv\hat{V}^{\dagger }\left\{ \alpha _{1}^{d}(t)+\alpha
_{2}^{d}Y_{u}^{2}+\alpha _{3}^{d}\text {Tr}(Y_{u}^{2})\right\} \hat{V},
\label{eqmvv}
\end{equation}
where the matrix $M^{{vv}}$ is introduced for notational simplicity.

The matrices $U_{L}^{d}\frac{d\left( U_{L}^{d}\right) ^{\dagger
}}{dt}$ and $\frac{d\left( U_{R}^{d}\right) }{dt}\left(
U_{R}^{d}\right) ^{\dagger }$ are 
antihermitian (this follows from unitarity) so their diagonal elements are
purely imaginary. The diagonal elements of the matrix $Y_{d}$ are purely
real and the matrix $M^{{vv}}$ is hermitian, so after taking the real part
of the diagonal elements of Eq.~(\ref{II.6}) we obtain 
\begin{equation}
\left( \frac{dY^{d}}{dt}\right) _{ii}=\frac{1}{(4\pi )^{2}}\left( M^{{vv}
}Y^{d}\right) _{ii}.  \label{II.7}
\end{equation}
The matrix $\left( M^{{vv}}\right) _{ij}$ has the simple form in our
approximation. 
\begin{equation}
\left( M^{{vv}}\right) _{ij} =\left( \hat{V}^{\dagger }\alpha
_{1}^{d}(t)\hat{V}+\hat{V}^{\dagger }\alpha
_{2}^{d}Y_{u}^{2}\hat{V}+\hat{V}^{\dagger }\alpha
_{3}^{d}\text {Tr}(Y_{u}^{2})\hat{V}\right) _{ij}
=\left\{ \alpha _{1}^{d}(t)+\alpha _{3}^{d}Y_{t}^{2}\right\} \delta
_{ij}+\alpha _{2}^{d}Y_{t}^{2}\hat{V}_{it}^{\dagger }\hat{V}_{tj}
+{\cal O}(\lambda^{4}).  \label{II.8}
\end{equation}
The last step in derivation of (\ref{II.8}) follows from the approximation
based on hierarchy 
\[
\left( Y^{u}\right) _{11}\thicksim \lambda ^{7}Y_{t},\quad \quad \left(
Y^{u}\right) _{22}\thicksim \lambda ^{3.5}Y_{t}. 
\]
Finally we obtain the following equations for the eigenvalues of the
\textit{down} quark Yukawa couplings 
\begin{equation}
 \frac{dY_{ii}^{d}}{dt}=\frac{1}{(4\pi )^{2}}\left\{ \alpha
_{1}^{d}+\alpha _{2}^{d}Y_{t}^{2}\left| \hat{V}_{ti}\right| ^{2}+\alpha
_{3}^{d}Y_{t}^{2}\right\} Y_{ii}^{d}+{\cal O}(\lambda^{4}).  \label{II.9}
\end{equation}
Using the hierarchical properties of the $\hat{V}$ matrix 
\begin{equation}
\left| \hat{V}_{td}\right| ^{2}\thicksim 
\lambda^{6},\,\,\,\,\,\,\,\,\,\,\,\left| 
\hat{V}_{ts}\right| ^{2}\thicksim \lambda ^{4},\,\,\,\,\,\,\,\,\left|
\hat{V}_{tb}\right| ^{2}=1,
\label{II.10}
\end{equation}
we obtain the following equations for the eigenvalues of the \textit{down} quarks
Yukawa couplings 
\begin{equation}
\begin{split}
\frac{dY_{d,s}}{dt} &=\frac{1}{(4\pi )^{2}}\left\{ \alpha _{1}^{d}+\alpha
_{3}^{d}Y_{t}^{2}\right\} Y_{d,s}
+{\cal O}(\lambda^{4}),  \\
\frac{dY_{b}}{dt} &=\frac{1}{(4\pi )^{2}}\left\{ \alpha _{1}^{d}+\left(
\alpha _{3}^{d}+\alpha _{2}^{d}\right) Y_{t}^{2}\right\} Y_{b}
+{\cal O}(\lambda^{4}).
\label{II.11}
\end{split}
\end{equation}
The solution of Eqs.~(\ref{II.11}) reads 
\begin{equation}
\begin{split}
Y_{d,s}(t) &=Y_{d,s}(t_{0})\sqrt{r^{\prime }(t)}h_{m}^{\alpha
_{3}^{d}}(t), \\
Y_{b}(t) &=Y_{b}(t_{0})\sqrt{r^{\prime }(t)}\,h_{m}^{\left( \alpha
_{2}^{d}+\alpha _{3}^{d}\right) }(t).  \label{II.12}
\end{split}
\end{equation}
where the function $r^{\prime}(t)$ is 
given in Eq.~(\ref{13}). These results complement previous results
shown in Sec.~\ref{section_II}.
Now, Eqs.~(\ref{II.12}) together with the VEV evolution Eq.~(\ref{21}) gives
the evolution of the \textit{down} quark masses 
\begin{equation}
m_{i}^{d}(t) =\frac{{v}_{d}(t)}{\sqrt{2}}Y_{i}^{d}(t)=\frac{{v}_{d}(t_{0})
}{\sqrt{2}}Y_{i}^{d}(t_{0})\sqrt{r^{\prime }(t)r^{\prime\prime}_{{v}_{d}}(t)}
h_{m}^{K_{i}^{d}}\left( t\right)
=m_{i}^{d}(t_{0})\sqrt{r^{\prime }(t)r^{\prime\prime}_{{v}_{d}}(t)}
h_{m}^{K_{i}^{d}}\left( t\right)  \label{II.13}
\end{equation}
there the powers $K_{i}^{d}$ are equal to 
\begin{equation}
K_{d,s}^{d}=\alpha _{3}^{d}+\alpha _{3}^{{v}_{d}},\,\,\,\,\,\,\,K_{b}^{d}=
\alpha _{2}^{d}+\alpha _{3}^{d}+\alpha _{3}^{{v}_{d}}.  \label{II.14}
\end{equation}
Eqs.~(\ref{II.13}) give the analytical form of the \textit{down} quark
mass evolution.
The evolution of the ratios of the \textit{down} quark masses for three
models SM, DHM, and MSSM is shown in Fig.~\ref{fig3}~(c) and Fig.~\ref{fig3}~(d).

\subsection{Cabibbo-Kobayashi-Maskawa matrix}
\label{section_III_B}

The other set of observables is related with the CKM~matrix. These
observables include the absolute values of the matrix elements of the
CKM~matrix, Jarlskog's parameter~$J$~\cite{n12a}, Wolfenstein
parameters $\lambda$, $A$, $\rho$, $\eta$~\cite{W,Buras2} and the
angles of the unitarity triangle~\cite{n12,Buras3}.

The evolution of the CKM~matrix is simpler than that for masses because the
CKM~matrix depends only on the left diagonalizing matrices of the biunitary
transformations for the Yukawa couplings and as was shown in
Ref.~\cite{ref0} the
evolution of CKM~matrix depends only on one function of the energy $h(t)$. This
fact implies that there exist correlations between the evolution of the
various elements of the CKM~matrix.

In the following we shall first find the evolution equations of the absolute
values of the CKM~matrix elements and afterwards analyze the $J$
parameter and the unitarity triangle.

\subsubsection{Evolution of the absolute values of the CKM~matrix $\hat{V}$.}
\label{section_III_B_1}

Our starting point is Eq.~(\ref{II.6}). We take now the imaginary part of
the diagonal elements and the full off-diagonal
elements of Eq.~(\ref{II.6}), and 
using the form of the matrix $M^{{vv}}$ in Eq.~(\ref{II.8}) we obtain 
\begin{equation}
\left( \left( U_{L}^{d}\right) \frac{d\left( U_{L}^{d}\right) ^{\dagger }}{dt
}\right) _{ii}+\left( \frac{d\left( U_{R}^{d}\right) }{dt}\left(
U_{R}^{d}\right) ^{\dagger }\right) _{ii}=0  \label{II.14X}
\end{equation}
and 
\begin{equation}
\left( \left( U_{L}^{d}\right) \frac{d\left( U_{L}^{d}\right) ^{\dagger }}{dt
}\right) _{ij}Y_{j}^{d}+Y_{i}^{d}\left( \frac{d\left( U_{R}^{d}\right) }{dt}
\left( U_{R}^{d}\right) ^{\dagger }\right) _{ij}=\frac{1}{(4\pi )^{2}}\alpha
_{2}^{d}Y_{t}^{2}\hat{V}_{ti}^{*}\hat{V}_{tj}Y_{j}^{d},
\;\;\;\;\;\; i\neq j.  \label{II.15}
\end{equation}
The hermitian conjugate of Eq.~(\ref{II.15}) reads 
\begin{equation}
Y_{i}^{d}\left( \left( U_{L}^{d}\right) \frac{d\left( U_{L}^{d}\right)
^{\dagger }}{dt}\right) _{ij}+\left( \frac{d\left( U_{R}^{d}\right) }{dt}
\left( U_{R}^{d}\right) ^{\dagger }\right) _{ij}Y_{j}^{d}=-\frac{1}{(4\pi
)^{2}}\alpha
_{2}^{d}Y_{t}^{2}Y_{i}^{d}\hat{V}_{ti}^{*}\hat{V}_{tj},
\,\,\,\,\,\,\,\,\,\,\,\,\,\,\,\,
\,\,\,\,\,\,\,i\neq j.  \label{II.16}
\end{equation}
Now multiplying Eq.~(\ref{II.15}) by $Y_{i}^{d}\,$ and
Eq.~(\ref{II.16}) by $Y_{j}^{d}\,$ and summing these equations, we
obtain after some simple manipulations the differential equation for
the evolution of the right diagonalizing matrix of the down sector
\begin{equation}
\left( \frac{d\left( U_{R}^{d}\right) }{dt}\left( U_{R}^{d}\right) ^{\dagger
}\right) _{ij}=-\frac{2Y_{i}^{d}Y_{j}^{d}}{\left( Y_{j}^{d}\right)
^{2}+\left( Y_{i}^{d}\right) ^{2}}\left( \left( U_{L}^{d}\right) \frac{
d\left( U_{L}^{d}\right) ^{\dagger }}{dt}\right) _{ij}.  \label{II.17}
\end{equation}
Eq.~(\ref{II.17}) gives the relation between the evolution of the left and
right diagonalizing matrices and it is the key relation that permits the
derivation of the evolution of the left diagonalizing matrix of the \textit{down}
quark Yukawa couplings. Inserting Eq.~(\ref{II.17}) in Eq.~(\ref{II.15}) we
obtain the evolution equation for the $U_{L}^{d}$ matrix 
\begin{equation}
\frac{\left( Y_{j}^{d}\right) ^{2}-\left( Y_{i}^{d}\right) ^{2}}{\left(
Y_{j}^{d}\right) ^{2}+\left( Y_{i}^{d}\right) ^{2}}\left( U_{L}^{d}\frac{
d\left( U_{L}^{d}\right) ^{\dagger }}{dt}\right) _{ij}=\frac{1}{(4\pi )^{2}}
\alpha
_{2}^{d}Y_{t}^{2}\hat{V}_{ti}^{*}\hat{V}_{tj}
\,\,\,\,\,\,\,\,\,\,\,\,\,\,\,\,\,\,\,\,\,i
\neq j.  \label{II.18}
\end{equation}
We will now convert Eq.~(\ref{II.18}) into equations for the CKM~matrix
elements. To this end we first notice that within our approximation
(neglecting all terms of the order $\lambda ^{4}$ and higher) 
\begin{equation}
\frac{\left( Y_{j}^{d}\right) ^{2}-\left( Y_{i}^{d}\right) ^{2}}
{\left(Y_{j}^{d}\right) ^{2}+\left( Y_{i}^{d}\right) ^{2}}
=-\text{sign}(i-j)+{\cal O}(\lambda^{4}),
\label{II.19}
\end{equation}
where $\text{sign}(i-j)=\mp 1$ for $i\lessgtr j$.

Next we know that the \textit{up} quark diagonalizing matrix $U_{L}^{u}$ does not
depend on the energy. Using the definition of the CKM~matrix
$\hat{V}=U_{L}^{u}\left( U_{L}^{d}\right) ^{\dagger }$ we obtain
\begin{equation}
\hat{V}^{\dagger }\frac{d\hat{V}}{dt}=\left( U_{L}^{d}\right) 
\left(U_{L}^{u}\right)
^{\dagger }U_{L}^{u}\frac{d\left( U_{L}^{d}\right) ^{\dagger }}{dt}=\left(
U_{L}^{d}\right) \frac{d\left( U_{L}^{d}\right) ^{\dagger }}{dt}.
\label{II.20}
\end{equation}
Now from Eqs.~(\ref{II.18}), (\ref{II.19}) and~(\ref{II.20}) the
following equations are obtained 
\begin{equation}
\left( \hat{V}^{\dagger }\frac{d\hat{V}}{dt}\right) _{ij}=-\frac{1}{(4\pi )^{2}}
\text{sign}(i-j)\alpha _{2}^{d}Y_{t}^{2}\hat{V}_{ti}^{*}\hat{V}_{tj}.
\label{II.21}
\end{equation}
Eq.~(\ref{II.21}) is valid only for $i\neq j$ but from the property that the
matrix $\left( \hat{V}^{\dagger }d\hat{V}/dt\right) $ is antiunitary
and its diagonal
elements are purely imaginary, we obtain equations for the absolute value of
the elements of the CKM~matrix 
\begin{subequations}
\label{II.22all}
\begin{eqnarray}
\frac{1}{\left|\hat{V}_{cd}\right|^2}\frac{d\left| \hat{V}_{cd}\right|
^{2}}{dt} &=&\frac{2}{(4\pi)^{2}}\alpha
_{2}^{d}Y_{t}^{2}\left| \hat{V}_{td}\right|^{2},
\label{II.22a} \\
\frac{1}{\left|\hat{V}_{td}\right| ^{2}}\frac{d\left|
\hat{V}_{td}\right| ^{2}}{dt}
&=&-\frac{2}{(4\pi )^{2}}\alpha
_{2}^{d}Y_{t}^{2}\left( 1-\left| \hat{V}_{td}\right| ^{2}\right),
\label{II.22b} \\ 
\frac{1}{\left| \hat{V}_{ub}\right|^{2}}\frac{d\left|
\hat{V}_{ub}\right| ^{2}}{dt}
&=&-\frac{2}{(4\pi )^{2}}\alpha
_{2}^{d}Y_{t}^{2}\left|
\hat{V}_{tb}\right| ^{2}, \label{II.22c} \\
\frac{1}{\left| \hat{V}_{cb}\right| ^{2}}\frac{d\left|
\hat{V}_{cb}\right| ^{2}}{dt}
&=&-\frac{2}{(4\pi )^{2}}\alpha
_{2}^{d}Y_{t}^{2}\left|
\hat{V}_{tb}\right| ^{2}, \label{II.22d}
\end{eqnarray}
and from the unitarity of the CKM~matrix we obtain the evolution of the
remaining CKM~matrix elements.

To solve Eqs.~(\ref{II.22all}) let us notice the following
relation
\end{subequations}
\label{II.23all}
\begin{equation}
\frac{2}{(4\pi )^{2}}\alpha _{2}^{d}Y_{t}^{2}=\frac{1}{h^{2}}\frac{dh^{2}}{dt
},  \label{II.23}
\end{equation}
where $h\left( t\right) \equiv\left[ h_{m}(t)\right] ^{\alpha
_{2}^{d}}$ 
and $h_{m}(t)$ is the
function introduced in Eq.~(\ref{7b}). Using relation (\ref{II.23}) we
obtain from Eq.~(\ref{II.22b})
\begin{subequations}
\label{II.24all}
\begin{equation}
|\hat{V}_{td}\left( t\right) |^{2}=\frac{|\hat{V}_{td}^{0}|^{2}}{h^{2}+\left(
1-h^{2}\right) |\hat{V}_{td}^{0}|^{2}},\;\;\;\;\;\;\;
\text{where}\,\,\,\,\hat{V}_{ij}^{0}\equiv\hat{V}_{ij}\left( t_{0}\right).
\label{II.24a}
\end{equation}
From Eqs.~(\ref{II.22a}) and ~(\ref{II.22b}) we obtain 
\begin{equation}
|\hat{V}_{cd}\left( t\right) |^{2}=\frac{h^{2}|\hat{V}_{cd}^{0}|^{2}}{
h^{2}+(1-h^{2})|\hat{V}_{td}^{0}|^{2}}.  \label{II.24b}
\end{equation}
Using the relation
$1-|\hat{V}_{tb}|^2=|\hat{V}_{ub}|^2+|\hat{V}_{cb}|^2$ we derive
from of Eqs.~(\ref{II.22c}) and~(\ref{II.22d}) the following
result\footnote{Note that the equation for the evolution  of
$|\hat{V}_{tb}|^2$ given in Ref.~\cite{ref0}, Eq.~(39d) has a missing
$h^2$ in the numerator.}
\begin{equation}
|\hat{V}_{tb}(t)|^{2}=
\frac{h^{2}|\hat{V}_{tb}^{0}|^{2}}{1+(h^{2}-1)|\hat{V}_{tb}^{0}|^{2}}.
\label{II.24c}
\end{equation}
and this result yields with the help of Eqs.~(\ref{II.22c}) and~(\ref{II.22d}
) the evolution of the $|\hat{V}_{ub}(t)|^{2}$ and $|\hat{V}_{cb}(t)|^{2}$:
\begin{equation}
|\hat{V}_{ub}\left( t\right) |^{2}=\frac{|\hat{V}_{ub}^{0}|^{2}}{
1+(h^{2}-1)|\hat{V}_{tb}^{0}|^{2}},  \label{II.24d}
\end{equation}
\begin{equation}
\left| \hat{V}_{cb}\left( t\right) \right| ^{2}=\frac{\left| 
\hat{V}_{cb}^{0}\right| ^{2}
}{1+\left( h^{2}-1\right) |\hat{V}_{tb}^{0}|^{2}}.  \label{II.24e}
\end{equation}
The remaining elements of the CKM~matrix are obtained from the
unitarity relation
\begin{equation}
\left| \hat{V}_{ud}\left( t\right) \right| ^{2}=\frac{h^{2}\left|
\hat{V}_{ud}^{0}\right| ^{2}}{h^{2}+\left( 1-h^{2}\right) 
\left| \hat{V}_{td}^{0}\right|
^{2}},  \label{II.24f}
\end{equation}
\begin{equation}
\left| \hat{V}_{us}\left( t\right) \right| ^{2}=
\left| \hat{V}_{us}^{0}\right|^{2}+
\left( h^{2}-1\right) 
\left[
\frac{|\hat{V}_{ub}^{0}|^{2}|\hat{V}_{tb}^{0}|^{2}}
{1+\left( h^{2}-1\right) |\hat{V}_{tb}^{0}|^{2}}
-\frac{|\hat{V}_{ud}^{0}|^{2}|\hat{V}_{td}^{0}|^{2}}
{h^{2}+\left(1-h^{2}\right) |\hat{V}_{td}^{0}|^{2}}
\right] ,  \label{II.24g}
\end{equation}
\begin{equation}
\left| \hat{V}_{cs}\left( t\right) \right| ^{2}=
\left| \hat{V}_{cs}^{0}\right|^{2}+
\left( h^{2}-1\right) 
\left[
\frac{|\hat{V}_{cb}^{0}|^{2}|\hat{V}_{tb}^{0}|^{2}}
{1+\left( h^{2}-1\right) |\hat{V}_{tb}^{0}|^{2}}
-\frac{|\hat{V}_{cd}^{0}|^{2}|\hat{V}_{td}^{0}|^{2}}
{h^{2}+\left(1-h^{2}\right) |\hat{V}_{td}^{0}|^{2}}
\right] ,  \label{II.24h}
\end{equation}
\begin{equation}
\left| \hat{V}_{ts}\left( t\right) \right| ^{2}=
\left| \hat{V}_{ts}^{0}\right|^{2}-
\left( h^{2}-1\right) 
\left[
\frac{|\hat{V}_{tb}^{0}|^{2}(1-|\hat{V}_{tb}^{0}|^{2})}
{1+\left( h^{2}-1\right) |\hat{V}_{tb}^{0}|^{2}}
-\frac{|\hat{V}_{td}^{0}|^{2}(1-|\hat{V}_{td}^{0}|^{2})}
{h^{2}+\left(1-h^{2}\right) |\hat{V}_{td}^{0}|^{2}}
\right].  \label{II.24i}
\end{equation}
\end{subequations}
Eqs.~(\ref{II.24all}) represent the RG evolution of the
CKM~matrix. The absolute values of the CKM~matrix elements
do not depend on the parameterization of the
CKM~matrix. Notice that in agreement with the Theorem~1 in
Ref.~\cite{ref0} the
evolution of the CKM~matrix depends only on one function $h(t)$
of energy. It is for this reason that we consider $h(t)$ as an universal
function of energy.

\subsubsection{Jarlskog Invariant}
\label{section_III_B_2}

Jarlskog parameter $J$ is defined as~\cite{n12a}
\begin{equation}
J=
\Im
\left[ \hat{V}_{ud}\hat{V}_{cs}\hat{V}_{us}^{*}\hat{V}_{cd}^{*}\right] =
\Im
\left[ \hat{V}_{ud}\hat{V}_{tb}\hat{V}_{ub}^{*}\hat{V}_{td}^{*}\right] =
\Im \left( D\right),  \label{60}
\end{equation}
and it is nonvanishing if the CKM~matrix is non CP invariant.

\begin{figure}[!tb]
\centering
\includegraphics{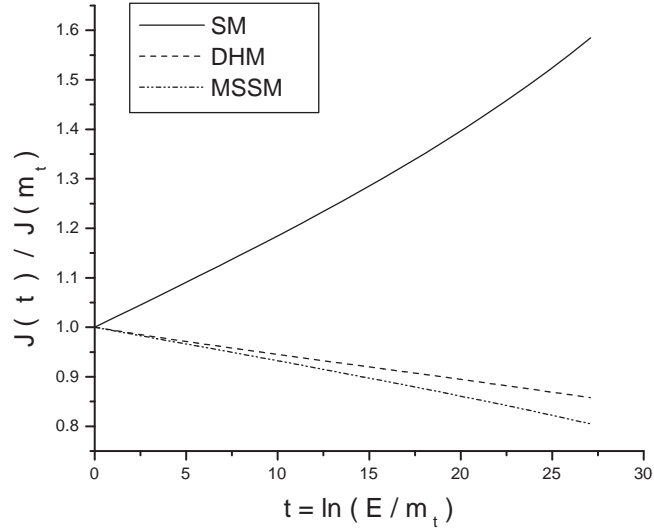}
\caption{\label{fig4}The scale dependence of the ratio
$J(t)/J(m_{t})$ for the Jarlskog invariant.}
\end{figure}

Let us now consider the following expression 
\begin{equation}
\frac{d\ln D}{dt}=\frac{d\ln \left( \hat{V}_{ud}\hat{V}_{tb}
\hat{V}_{ub}^{*}\hat{V}_{td}^{*}\right) 
}{dt}=\frac{1}{\hat{V}_{ud}}\frac{d}{dt}\hat{V}_{ud}+\frac{1}
{\hat{V}_{tb}}\frac{d}{dt}\hat{V}_{tb}+
\frac{1}{\hat{V}_{ub}^{*}}\frac{d}{dt}\hat{V}_{ub}^{*}
+\frac{1}{\hat{V}_{td}^{*}}\frac{d}{dt}
\hat{V}_{td}^{*}.  \label{61}
\end{equation}
Now using Eq.~(\ref{II.21}) and the property that the matrix $\hat{V}^{\dagger }
\frac{d\hat{V}}{dt}$ is antihermitian (and has imaginary diagonal elements) we
obtain 
\begin{equation}
\frac{d\ln D}{dt}=-2\alpha _{2}^{d}Y_{t}^{2}\left( \left| \hat{V}_{tb}\right|
^{2}-\left| \hat{V}_{td}\right| ^{2}\right) \approx -2\alpha
_{2}^{d}Y_{t}^{2}\left| \hat{V}_{tb}\right| ^{2},  \label{63}
\end{equation}
and thus the Jarlskog invariant fulfills the equation 
\begin{equation}
\frac{d\ln J}{dt}=-2\alpha _{2}^{d}Y_{t}^{2}\left| 
\hat{V}_{tb}\right|^{2}=-\frac{
1}{h^{2}}\frac{dh^{2}}{dt}\left| \hat{V}_{tb}\right| ^{2}  \label{64}
\end{equation}
which gives using Eq.~(\ref{II.24c}) 
\begin{equation}
\ln \frac{J}{J_{0}}=\ln \frac{1}{\left|
1+(h^{2}-1)|\hat{V}_{tb}^{0}|^{2}\right| }
\label{65}
\end{equation}
or 
\begin{equation}
J=\frac{J_{0}}{\left| 1+(h^{2}-1)|\hat{V}_{tb}^{0}|^{2}\right|
}\approx \frac{J_{0}
}{h^{2}}.  \label{66}
\end{equation}
The evolution of the Jarlskog invariant is given in Fig.~\ref{fig4}.

\subsubsection{The Wolfenstein Parameters}
\label{section_III_B_3}

Eqs.~(\ref{II.24all}) can be written in an approximate form
using the hierarchy of the CKM~matrix, given by the Wolfenstein
parameterization~\cite{W,Buras2}
\begin{equation}
\hat{V}=\left( 
\begin{array}{ccc}
1-\frac{1}{2}\lambda ^{2} & \lambda & A\lambda ^{3}(\rho -i\eta ) \\ 
-\lambda & 1-\frac{1}{2}\lambda ^{2} & A\lambda ^{2} \\ 
A\lambda ^{3}(1-\bar{\rho} -i\bar{\eta} ) & -A\lambda ^{2} & 1
\end{array}
\right),\;\;\;
\bar{\rho}=\rho\left(1-\frac{\lambda^2}{2}\right),\;\;\;
\bar{\eta}=\eta\left(1-\frac{\lambda^2}{2}\right).
  \label{67}
\end{equation}
Now neglecting all the terms of the relative order $\lambda ^{4}$ and
higher, we obtain
\begin{eqnarray}
&\displaystyle
1-\left| \hat{V}_{ud}\right| ^{2}
\approx
1-\left| \hat{V}_{ud}^{0}\right| ^{2},\;\;\;\;\;
\left| \hat{V}_{us}\right| ^{2} \approx \left| 
\hat{V}_{us}^{0}\right|^{2},\;\;\;\;\;
\left| \hat{V}_{ub}\right| ^{2} \approx \frac{\left| 
\hat{V}_{ub}^{0}\right|^{2}}{h^{2}} \nonumber\\
&\displaystyle
\left| 
\hat{V}_{cd}\right| ^{2}\approx \left| \hat{V}_{cd}^{0}\right| ^{2},
\;\;\;\;\;
1-\left| \hat{V}_{cs}\right| ^{2} \approx 1-\left| 
\hat{V}_{cs}^{0}\right|^{2}
+\frac{1-h^2}{h^2}\left|\hat{V}_{cb}^{0}\right|^{2},\;\;\;\;\;
\left| \hat{V}_{cb}\right|^{2}\approx \frac{\left|
\hat{V}_{cb}^{0}\right| ^{2}}{h^{2}},  \label{68} \\
&\displaystyle
\left| \hat{V}_{td}\right| ^{2} \approx 
\frac{\left| \hat{V}_{td}^{0}\right| ^{2}}{h^{2}},\;\;\;\;\;
\left| \hat{V}_{ts}\right| ^{2}\approx 
\frac{\left|\hat{V}_{ts}^{0}\right| ^{2}}{h^{2}},\;\;\;\;\;
1-\left| \hat{V}_{tb}\right| ^{2}\approx \frac{1-\left|
\hat{V}_{tb}^{0}\right| ^{2}}{h^2}.  \nonumber
\end{eqnarray}
From Eqs.~(\ref{68}) immediately follows the simple evolution of the
Wolfenstein parameters
\begin{equation}
A\left( t\right) =\frac{A}{h(t)},\,\,\,\,\,\,\,\,\,\,\text{and\thinspace
\thinspace \thinspace \thinspace \thinspace \thinspace \thinspace }\lambda
,\rho ,\eta \text{ are invariant.}  \label{69}
\end{equation}
Notice that the dependence of the CKM matrix and Wolfenstein
parameters on the renormalization scheme is given in~Ref.~\cite{kniehl}. 

\subsubsection{The Unitarity triangle}
\label{section_III_B_4}

The unitarity triangle~\cite{n12,Buras3} is obtained from the scalar product
of the first column of the $\hat{V}$ matrix by the complex conjugate of the
third
\begin{equation}
\hat{V}_{ud}\hat{V}_{ub}^{*}+\hat{V}_{cd}\hat{V}_{cb}^{*}
+\hat{V}_{td}\hat{V}_{tb}^{*}=0  \label{70}
\end{equation}
and then by rescaling it in such a way that the lengths of the sides
of the resulting triangle are equal
\begin{equation}
R_b=(1-\frac{\lambda^2}{2})\frac{1}{\lambda}
\left|\frac{\hat{V}_{ub}}{\hat{V}_{cb}}\right|,\;\;\;\;\;
R_t=\frac{1}{\lambda}
\left|\frac{\hat{V}_{td}}{\hat{V}_{cb}}\right|,\;\;\;\;\;1.
\end{equation}
From Eqs.~(\ref{68}) and~(\ref{69}) it immediately follows that $R_b$
and $R_t$ are invariant upon the evolution which implies that
the unitarity triangle is also invariant. Thus the complex phases of the
CKM matrix elements $\hat{V}_{td}$ and $\hat{V}_{ub}$ (angles $\beta$
and $\gamma$) are also invariant up to the order $\lambda^{4}$.

\section{Conclusions}
\label{section_IV}
In this paper we analyzed the solutions of the RGE for the quark
Yukawa couplings, for the Higgs VEV's and also the evolution of all the
observables that follow from them. The results depend on the model
dependent functions given in Eq.~(\ref{22}). The most interesting is
the universal function $h(t)$ because only on this function depends the
evolution of the CKM matrix, Eq.~(\ref{II.24all}) or~(\ref{68}). The
running of the absolute values of 
the CKM matrix elements and the invariance of the unitarity triangle
angles implies that the evolution of the CKM matrix is remarkably simple
\begin{equation}
\left(
\begin{array}{ccc}
\displaystyle\hat{V}_{ud}^{0}
&\displaystyle\hat{V}_{ud}^{0}
&\displaystyle\hat{V}_{ub}^{0}\\
\displaystyle\hat{V}_{cd}^{0}
&\displaystyle\hat{V}_{cs}^{0}
&\displaystyle\hat{V}_{cb}^{0}\\
\displaystyle\hat{V}_{td}^{0}
&\displaystyle\hat{V}_{ts}^{0}
&\displaystyle\hat{V}_{tb}^{0}
\end{array}
\right)
\longrightarrow
\left(
\begin{array}{ccc}
\displaystyle\hat{V}_{ud}^{0}
&\displaystyle\hat{V}_{ud}^{0}
&\displaystyle\frac{\hat{V}_{ub}^{0}}{h}\\
\displaystyle\hat{V}_{cd}^{0}
&\displaystyle\hat{V}_{cs}^{0}
&\displaystyle\frac{\hat{V}_{cb}^{0}}{h}\\
\displaystyle\frac{\hat{V}_{td}^{0}}{h}
&\displaystyle\frac{\hat{V}_{ts}^{0}}{h}
&\displaystyle\hat{V}_{tb}^{0}
\end{array}
\right)
\label{evCKM}
\end{equation}
We observe (see Fig.~\ref{fig2}) that the function $h(t)$ is
decreasing for the SM and increasing for the DHM and MSSM. This
results in a qualitatively different evolution of the matrix elements
and the observables related to the CKM matrix for the SM in comparison
to the DHM and MSSM. An important result is
the dependence on it of the Jarlskog invariant which is shown in
Fig.~\ref{fig4}. We thus see that the CP violation is enhanced with
increasing energy in the SM while it decreases for the DHM and MSSM.

The evolution of the unitarity triangle is also very important. We
have shown that the angles of the unitarity triangle remain constant
upon the evolution. The invariance of the angles of the unitarity
triangle is very significant because it means that at the grand
unification scale the angles are the same as at the low energy so if
there is a symmetry at the grand unification scale then it has to
predict the low energy angles of the unitarity triangle. This strongly
constrains the possible symmetries or textures at the grand
unification scale.

As we discussed in Section~\ref{section_I} our results are approximate
and we kept the terms up to the order $\lambda^{3}$. In
the next order one has to keep the powers up to the order
$\lambda^{4}$. In this approximation the results are qualitatively the
same and the only corrections are small modifications of the functions
$h_{m}(t)$ and $h(t)$. The next order, $\lambda^{5}$, is significantly
different and it will be discussed elsewhere.

\begin{acknowledgments}
We gratefully acknowledge the financial support from
CONACYT--Proyecto~ICM (Mexico). S.R.J.W.  also thanks to
``Comisi\'{o}n de Operaci\'{o}n y Fomento de Actividades
Acad\'{e}micas'' (COFAA) from Instituto Polit\'{e}cnico Nacional.
\end{acknowledgments}


\begin{thebibliography}{}




\bibitem{ref1}  R.D.~Peccei and K.~Wang, Phys.\ Rev.\ \textbf{D53}, 2712
(1996); H.~Gonz\'{a}lez et al., Phys.\ Lett.\ \textbf{B 440}, 94 (1998);
H.~Gonz\'{a}lez et al., \emph{A New symmetry of Quark Yukawa Couplings}, page
755, International Europhysics Conference on High Energy Physics, Jerusalem
1997, eds.\ Daniel Lellouch, Giora Mikenberg, Eliezer Rabinovici,
Springer-Verlag 1999.

\bibitem{ref2}  K.S.~Babu, Z.~Phys.~\textbf{ C~35}, 69 (1987); P.~Binetruy and
P.~Ramond, Phys.\ Lett.\ \textbf{ B350}, 49 (1995); K.~Wang, Phys.\
Rev.\ \textbf{D54}, 5750 (1996).

\bibitem{ref3}  B.~Grzadkowski and M.~Lindner, Phys.\ Lett.\ \textbf{ B193}, 71
(1987); B.~Grzadkowski, M.~Lindner and S.~Theisen, Phys.\ Lett. \textbf{ B198},
64 (1987); M.~Olechowski and S.~Pokorski, Phys.\ Lett.\ \textbf{ B 257}, 388
(1991); G.~Cvetic, C.S.~Kim and S.S.~Hwang, Int.\ J.\ Mod.\ Phys.\ \textbf{ A14}
769 (1999).M.E.~Machacek and M.T.~Vaughn, Nucl.\ Phys.\ \textbf{ B222}, 83
(1983); \textbf{ B236}, 221 (1984); \textbf{ B249}, 70 (1985);
C.~Balzereit, Th.~Hansmann, T.~Mannel and B.~Pl\"umper, Eur.\ Phys.\
J. \textbf{C9} 197 (1999), \texttt{hep-ph/9810350}.

\bibitem{ref4}  H.~Arason, et al, Phys.\ Rev.\ \textbf{ D46}, 3945
(1992);
H.~Arason, et al, Phys.\ Rev.\ \textbf{ D47}, 232 (1992);
D.J.~Casta\~no, E.J.~Piard and P.~Ramond, Phys.\ Rev.\ \textbf{ D49},
4882 (1994).

\bibitem{ref7}  H.~Fusaoka and Y.~Koide, Phys.\ Rev.\ \textbf{ D57}, 3986 (1998).

\bibitem{ref5}  K.~Sasaki, Z.\ Phys.\ \textbf{ C~3}2, 149 (1986).

\bibitem{n2}  S.R.~Ju\'{a}rez~W., S.F.~Herrera, P.~Kielanowski and G.~Mora,
\textit{Energy dependence of the quark masses and mixings} in
Particles and Fields,
Ninth Mexican School, Metepec, Puebla, M\'{e}xico. AIP Conference
Proceedings 562 (2001) 303-308, ISBN 1-56396-998-X, ISSN 0094-243X,
\texttt{hep-ph/0009148}.

\bibitem{n3}  N.~Nimai Singh, Eur.\ Phys.\ J.\ \textbf{C19}, 137
(2001), \texttt{hep-ph/0009211}.

\bibitem{n4}  Yoshio Koide, Hideo Fusaoka, Phys.\ Rev.\ \textbf{D64},
053014  (2001), \texttt{hep-ph/0011070}.

\bibitem{n6}  C.R.\ Das, M.K.\ Parida, Eur.\ Phys.J.\ \textbf{C20}, 121 (2001).
\texttt{hep-ph/0010004}.

\bibitem{n7}  Cheng-Wei Chiang, Phys.\ Rev.\ \textbf{D63}, 076009
(2001), \texttt{hep-ph/0011195}.

\bibitem{n9}  N.~Cabibbo, Phys.\ Rev.\ Lett.\ \textbf{10}, 531 (1963).

\bibitem{n10}  M.~Kobayashi, T.~Maskawa, Prog.\ Theor.\
Phys.\ \textbf{49}, 652 (1973). 

\bibitem{ref0}  P.~Kielanowski, S.R.~Ju\'{a}rez~W.\ and G.~Mora, 
Phys.\ Lett.\ \textbf{B479}, 181 (2000). S. R.~Ju\'{a}rez~W., P.~Kielanowski and
G.~Mora, \emph{New Properties of the Renormalization Group Equations of the
Yukawa Couplings and CKM~Matrix} in the Proceedings of the Eighth Mexican
School \emph{Particles and Fields}, Oaxaca, M\'{e}xico. AIP Conference
Proceedings 490 (1998), 351-354, ed.\ J.C.~D'Olivo, G.~L\'{o}pez C.\ and
M.~Mondrag\'{o}n, L.C.~Catalog Card No.\ 99-067150, ISBN 1-56396-895-9, ISSN
0094-243X., DOE CONF-981188.

\bibitem{ref0A} Notice that in Ref.~\cite{ref0} we used a slightly
unfortunate notation by denoting the eigenvalues of the QYC
matrices by the symbol $m_i$ which coincided with the notation for the
quark masses. In this paper we use the symbol $Y_i$ for the
eigenvalues of the QYC matrices and $m_i$ for the quark masses.

\bibitem{PDG}
K.~Hagiwara {\it et al.}, Phys.\ Rev.\ {\bf D66}, 010001 (2002).

\bibitem{n12a}
C.~Jarlskog,  Phys.\ Rev.\ Lett.\ \textbf{ 55} 1039 (1985) and Zeit.\
f.\ Phys.\ \textbf{C29}, 491 (1985).

\bibitem{W}  L.~Wolfenstein, Phys.\ Rev.\ Lett.\ \textbf{51}, 1945 (1983).

\bibitem{Buras2}
A.J.~Buras, M.E.~Lautenbacher and G.~Ostermaier, Phys.\ Rev.\
\textbf{D 50}, 3433 (1994).

\bibitem{kniehl}
B.A.~Kniehl, F.~Madricardo and M.~Steinhauser,
Phys.~Rev. \textbf{D 62}, 073010 (2000).

\bibitem{n12}  C.~Jarlskog and R.~Stora, Phys.\ Lett.\ \textbf{ B208},
268 (1988); L.-L.~Chau and W.-Y.~Keung,
Phys.\ Rev.\ Lett.\ \textbf{53}, 1802 (1984).

\bibitem{Buras3} Andrzej J.~Buras,
\emph{Flavour Physics and CP Violation in the SM},
Introductory Lecture given at KAON 2001, Pisa, 12 June--17 June, 2001,
\texttt{hep-ph/0109197}.
\end{thebibliography}
\end{document}